\documentclass[pra,onecolumn,eqsecnum,superscriptaddress]{revtex4}

\usepackage{amssymb, amsbsy, amsmath, latexsym, dsfont, array, layout, graphicx}
\usepackage{color}

\newcommand{\ket}[1]{\left|{#1}\right\rangle}
\newcommand{\bra}[1]{\left\langle{#1}\right|}

\begin{document}

\title{Bipartite entanglement purification with neutral atoms}
\author{P. Xue}
\affiliation{Institute for Quantum Information Science, University of Calgary, Canada T2N 1N4}
\author{X.-F. Zhou}
\affiliation{Key Laboratory of Quantum Information, University of Science and Technology of
China, Hefei 230026, China}
\date{\today}

\begin{abstract}
We theoretically study bipartite entanglement purification with neutral atoms via cavity-assistant interaction and linear
optical elements. We focus on entanglement distillation and the recurrence protocol, whose performances under
idealized and realistic conditions are discussed. The implementation of these purification protocols has been tested with
numerical simulations. We analyze the performance and stability of all required operations and emphasize that all techniques are feasible with current experimental technology.
\end{abstract}

\pacs{03.67.Mn, 03.67.Hk, 42.50.-p}
\maketitle

\section{introduction}

Entanglement has always been at the heart of quantum mechanics and plays a key role in quantum information science. Most
applications of quantum information processing require maximally entangled states. Many efforts have been undertaken to
generate entangled states in different physical systems. Recently, the controlled manipulation of certain systems in such a
way that entangled states can be generated on demand has become possible. However, noise in such control operations as well as
interactions with an uncontrollable environment result that the desired entangled states are created only with a
certain and less than unit fidelity. There are several ways to protect quantum information and entangled quantum states in
particular including quantum error correction \cite{error correction1, error correction2, error correction3, error correction4, error correction5,error correction6,error correction7,error correction8,error correction9} and entanglement purification which have been derived in
\cite{purification1,purification2,purification3} and experimentally demonstrated in \cite{experiment,Kwiat}. Here we will focus on the latter method and
the physical implementation with neutral atoms.

A number of different entanglement purification protocols exist, which differ in their purification range, the efficiency, and
the number of copies of the states they operate on. In the following, we show the physical implementation of entanglement
purification on neutral atoms via cavity-assistant interaction and linear optical elements. We demonstrate the distillation of
maximally entangled states from both non-maximally and mixed entangled inputs. In Sec.~II, we implement a positive operator
valued measure (POVM) and full Bell state measurement (BSM) on atoms via cavity-assistant interaction and linear optical
elements, which are basic tools for entanglement purification. In Sec.~III, we will consider bipartite entanglement
distillation (also called as filtering protocols) which operates on a single copy, as well as the recurrence protocol which
operates on two copies simultaneously at each step in Sec.~IV. With linear optical elements, we perform a filtering process to
maximize the entanglement of pure atomic entangled states. We have also applied our methods to initial states which are
partially mixed. After filtering, the distilled states demonstrate certain non-local correlations, as evidenced by their
violation of a form of Bell's inequality. Because the initial states do not have this property, they can be said to possess
``hidden" non-locality. Furthermore, we propose the physical implementation of the recurrence protocol with the same setup. In
Sec.~V, the implementation of the proposals have been tested with numerical simulations, which included various sources of
noise present in experiments, and was found to be robust against the influence of noise. We analyze the feasibility of the
implementation of these purification protocols and found it practically based on the POVM and BSM with high fidelities in the
current status of experimental technology. Remarkably, our protocol will enable the first-ever implementation of atomic
entanglement distillation and purification.

\section{theoretical model and basics tools}
\subsection{Theoretical Model}
\begin{figure}[ht]
   \includegraphics[width=.45\textwidth]{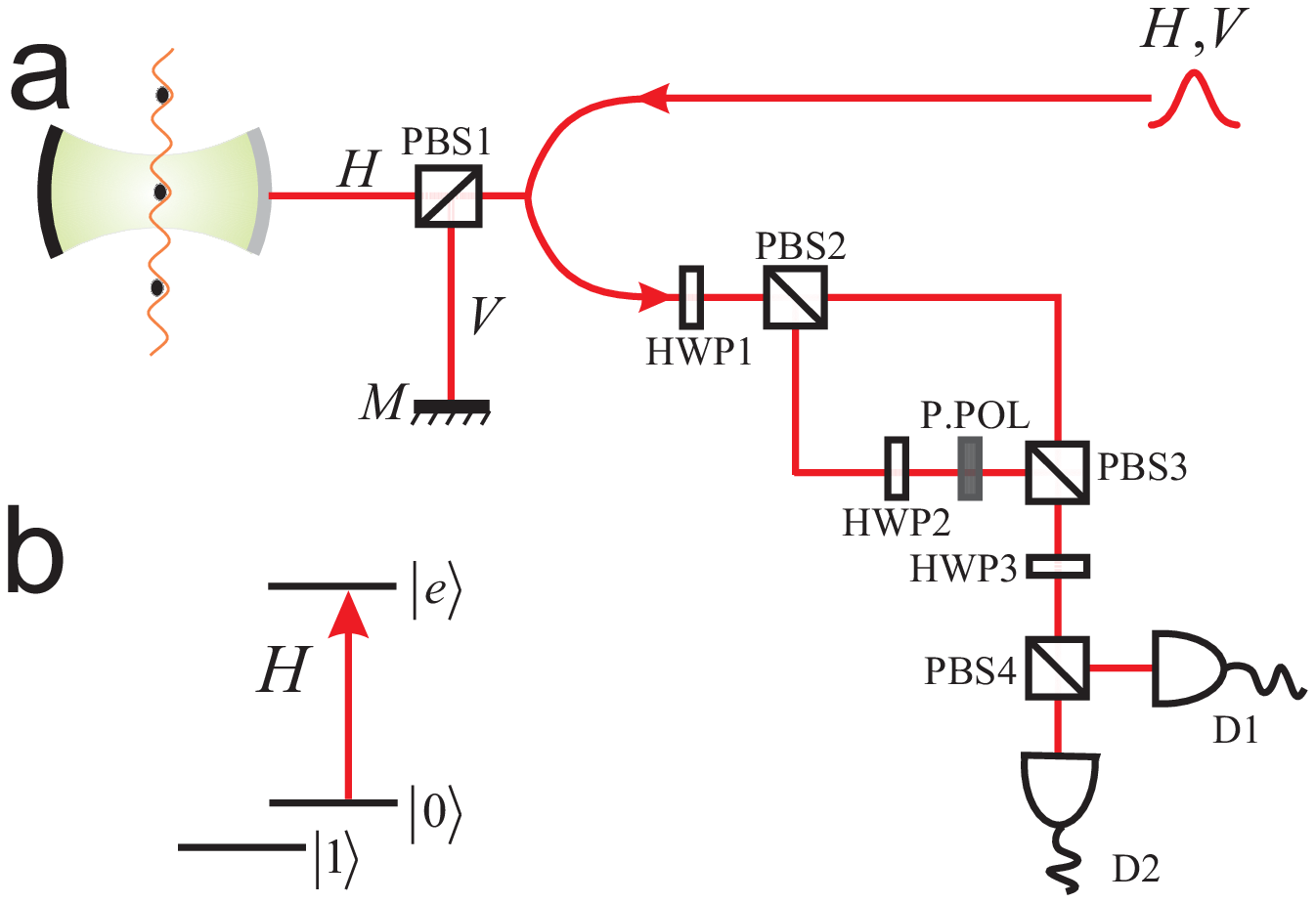}
   \caption{(a) Schematic setup for the implementation of the POVM on atoms through photon-scattering. Moving the optical lattice takes single atoms in and out of the cavity. Scattered photon pulses leak out of the cavity after a reflection, pass throught the linear optical elements and are finally detected by two single-photon detectors. For a detection event in $D1$, the outcome $P_{1}$ is obtained directly; for a detection event in $D2$ we obtain $P_{1}$ after a single-qubit rotation applied to the atom. (b) Relevant energy level structure of the atoms and their coupling configuration.}
   \label{fig1}
\end{figure}

Our source of entanglement and some basics tools which are used to implement entanglement purification include a POVM and BSM,
which are taken as cavity-assistant interaction on neutral atoms. The atoms are trapped in a transverse optical lattice and
thus can be taken into and out of the cavity for local operations in order to circumvent the requirement for individual
addressing. We consider a high Q optical cavity couples photons to the certain transitions in atoms whose internal states are
shown in Fig.\ \ref{fig1}b. The key parameters describing a cavity QED system are the cavity resonance frequency $\omega_\text{c}$, the
frequency $\omega_\text{a}=(\omega_\text{c})$ of the atomic transition from the ground state $\ket{0}$ to the excited state
$\ket{e}$. The other ground state $\ket{1}$ is decoupled due to the large hyperfine splitting. The system of the cavity and
atoms are described by the Hamiltonian (setting $\hbar=1$)
\begin{equation}
H=H_\text{o}+H_\text{int}+H_\gamma+H_\kappa ,
\end{equation}
where
\begin{align}
H_\text{o}=\omega_\text{c}a^\dagger a + \sum_{j}\omega_\text{a} (\ket{e}_j\bra{e}-\ket{0}_j\bra{0}); \notag \\
H_\text{int}=\sum_{j}g_j\left(a_c\ket{e}_j\bra{0}+a_c^\dagger\ket{0}_j\bra{e}\right),
\end{align}
and $H_\kappa$ describes the coupling of the cavity to the continuum which produces the cavity decay rate
$\kappa=\omega_\text{c}/Q$, while $H_\gamma$ describes the coupling of the atom to modes other than the cavity mode which
cause the excited state to decay at rate $\gamma$.

We use the single-photon pulse resonant with the bare cavity mode with the horizontal polarization h. Thus the input state of
cavity is $\ket{h}$. Suppose we perform this operation in the limit with $T\gg 1/\kappa$, that is the pulse duration T is much
large than the time that the photon state in the cavity. We now present a detailed theoretical model with one atom as an
example. In the rotating wave approximation, the Hamiltonian of atom-cavity and free space is
\begin{align}
H_{\mathrm{sy}}=&-\mathrm{i}\frac{\gamma}{2}\ket{e} \bra{e}
+g(a_c\ket{e}\bra{0}+\mathrm{h.c.})+\Delta a_c^\dagger a_{c} \notag\\
&+\int_{-\infty}^{\infty}\omega d\omega b^{\dagger}\left(
\omega\right)b\left(\omega\right) \\
&+\mathrm{i}\sqrt{\frac{\kappa}{2\pi}}\int_{-\infty}^{\infty}d\omega\left[a_{c}b^{\dagger}\left(\omega\right)-a_c^\dagger
b\left(\omega\right)\right], \notag
\end{align}
where $\Delta$ denotes the detuning of the cavity field mode $a_{c}$ from the atomic transition, and $b\left(\omega\right)$
with the standard relation $\left[b\left(\omega\right),b^{\dagger}\left(
\omega'\right)\right]=\delta\left(\omega-\omega'\right)$ denotes the one-dimensional free-space modes which couple to the
cavity mode $a_{c}$. According to the quantum Langevin equation and the boundary condition of the cavity, we can deduce that
the single-sided cavity input and output field operators $b_{\mathrm{in}}\left( t\right)$ and $b_{\mathrm{out}}(t)$ are
connected with the cavity mode $a_c(t)$ through the relations \cite{Wal1,Wal2}
\begin{equation}
\dot{a}_c(t)=-\mathrm{i}\left[a_c(t),H_\text{sy}\right] -\left(\mathrm{i}\Delta+\frac{\kappa}{2}\right)a_c(t)
-\sqrt{\kappa}b_{\mathrm{in}}(t)
\end{equation}
and
\begin{equation}
b_{\mathrm{out}}(t)=b_{\mathrm{in}}(t)+\sqrt{\kappa}a_{c}(t),
\end{equation}
where the Hamiltonian $H_\text{sy}$ shown in Eq.~(2.3) describes the coherent interaction between the atom and the cavity mode
$a_{c}$. The operators $b_{\mathrm{in}}(t)$ and $b_{\mathrm{out}}(t)$ satisfy the following commutation relations
$\left[b_{\mathrm{in}}(t),b_{\mathrm{in}}^\dagger\left(t'\right)\right]
=\delta\left(t-t'\right)=\left[b_{\mathrm{out}}(t),b_{\mathrm{out}}^\dagger\left(t'\right)\right]$. If the atom is in the
state $\ket{1}$, the Hamiltonian $H_\text{sy}$ is not active, and induces $\Delta=0$. When the input pulse shape changes
slowly with time $t$ compared with the cavity decay rate $\kappa$, from Eqs.~(2.4) and (2.5), we obtain
$b_{\mathrm{out}}(t)\approx\frac{\mathrm{i}\Delta-\kappa/2}{\mathrm{i}\Delta+\kappa/2}b_{\mathrm{in}}(t)=-b_{\mathrm{in}}(t)$.
While, if the atom is in the state $\ket{0}$, in the case of strong coupling \cite{Aga1,Aga2,Aga3}, the effective detunings of two
dressed cavity modes from the input pulse are $\Delta=\pm g$. Consider that the system works in the strong coupling regime
$g\gg\kappa$, we obtain $b_{\mathrm{out}}(t)\approx b_{\mathrm{in}}(t)$. In practice it turned out that the result is true
even if $g\thicksim\kappa$ \cite{Duan2}. From the description above, we conclude that the state of the whole system of
atom-cavity and free-space acquires the phase $\pi$ or $0$, for $b_{\mathrm{out}}(t)\approx-b_{\mathrm{in}}(t)$ or
$b_{\mathrm{out}}(t)\approx b_{\mathrm{in}}(t)$, after the photon pulse has been reflected by the cavity. The input-output
process can be characterized by
\begin{equation}
e^{{\mathrm{i}\pi\ket{1,h}\bra{1,h}}}(a\ket{0}+b\ket{1})\ket{h}=(a\ket{0}-b\ket{1})\ket{h},
\end{equation} where we have discarded the
state of cavity since in the limit $T\gg 1/\kappa$, it is always in the vacuum state.

This model has been used to implement quantum gate and entanglement generation \cite{Duan2,Duan11,Duan12,Duan13,Xue}. In our paper we will introduce
for the first time how to use this model to implement a POVM on atomic state which is not trivial and very useful to quantum
information processing such as entanglement purification.

\subsection{Physical Implementation of a POVM}

According to Fig.\ \ref{fig1}a, if we choose
\begin{equation}
(a\ket{0}+b\ket{1})\left(\ket{h}+\ket{v}\right)/\sqrt{2}
\end{equation}
as initial state of the atom+photon system, then the photons with horizontal and vertical polarizations are ``bounced" back
from the atom-cavity system and a mirror M respectively. After leaving the cavity and passing through the first polarizing
beam splitter (PBS1), which, as we recall, reflects vertically polarized light $\ket{v}$ and transmits horizontally polarized
light $\ket{h}$, the polarization of the photon is rotated by $\pi/4$ using half wave plate 1 (HWP1) which performs
$\ket{h}\rightarrow\left(\ket{h}+\ket{v}\right)/\sqrt{2}$ and $\ket{v}\rightarrow \left(\ket{h}-\ket{v}\right)/\sqrt{2}$, and
the state
\begin{equation}
a\ket{0}\ket{h}+b\ket{1}\ket{v}
\end{equation} is obtained. Afterwards, the photon passes through PBS2. The polarization of the reflected photon is subsequently
rotated by HWP2. The extra rotation allows the applied partial polarizer to operate in the basis
$\left\{\ket{h},\ket{v}\right\}$ as
\begin{equation}
\ket{h}\rightarrow\varepsilon\ket{h}, \ket{v}\rightarrow\ket{v}.
\end{equation}
This filtered state then interferes at PBS3 with the state that has been transmitted at PBS2. The photon passes HWP3 and PBS4
and is then detected by single-photon detectors. If detector D2 clicks, the resulting atomic state is $a\ket{0}+ \varepsilon
b\ket{1}$ (not normalized). That is we implement a POVM on the atomic state as
\begin{equation}
P=\ket{0}\bra{0}+\varepsilon\ket{1}\bra{1}.
\end{equation}

The partial polarizer \cite{Zeilinger} can be realized experimentally by inserting into one path a series of slabs, tilted
about the vertical axis by $\theta$ (see Appendix for details). Owing to the well known polarization dependent reflectivity,
the transmitted photons are preferentially vertically polarized. In the ideal case, the vertical polarized photons are
perfectly transmitted, while the horizontal polarized ones are partially reflected. The transmission probability for
horizontal polarization is $T_\text{h}=\varepsilon$, while for vertical polarization it is only $T_\text{v}=1$.

In addition, if one adds a quarter wave plate (QWP) between partial polarizer and PBS3, another POVM
\begin{equation}
P'=\ket{0}\bra{0}+\mathrm{i}\varepsilon\ket{1}\bra{1}
\end{equation} can be easily implemented on atomic states.

\subsection{Physical Implementation of full BSM}

For two atoms which are trapped in different optical lattices and both interacting with the cavity, we can implement a BSM on
two atoms, which has been mentioned briefly in our recent work \cite{Xue}. Here we give more detailed description of BSM on
atoms.

First the single photon pulse enters the cavity with only atom 1 trapped in optical lattice A. After the interaction between
atom and cavity mode, a gate operation $e^{\mathrm{i}\pi\ket{1,h}\bra{1,h}}$ is applied on the atom and the photon pulse. Atom
2 trapped in optical lattice B now is moved into the cavity while atom 1 is outside, and the pulse is reflected successively
to enter the cavity again, so that the same operation is applied on atom 2 and the pulse. That is
\begin{widetext}
\begin{align}
&\prod_{i=1,2} e^{\mathrm{i}\pi\ket{1}_i\bra{1}\otimes\ket{h}\bra{h}}(a_1\ket{0}_1+b_1\ket{1}_1)
(a_2\ket{0}_2+b_2\ket{1}_2)(\ket{h}+\ket{v})/\sqrt{2}\notag\\
&=(a_1\ket{0}_1-b_1\ket{1}_1) (a_2\ket{0}_2-b_2\ket{1}_2)\ket{h}+(a_1\ket{0}_1+b_1\ket{1}_1)
(a_2\ket{0}_2+b_2\ket{1}_2)\ket{v}.
\end{align}
\end{widetext}
Finally, the photon passes through a HWP and is detected by either D1 or D2, we obtain the projection if the detectors D1 and
D2 click respectively
\begin{equation}
P_{1}=\left\vert 00\right\rangle \left\langle 00\right\vert + \left\vert 11\right\rangle \left\langle 11\right\vert;
P_{2}=\openone-P_{1}.
\end{equation}

Performing the measurement $\left\{  P_{1},P_{2}\right\}  $ on atoms allows one to distinguish the subspace spanned by
$\left\{  \left\vert \Phi^{+}\right\rangle ,\left\vert \Phi^{-}\right\rangle \right\}  $ and $\left\{  \left\vert
\Psi^{+}\right\rangle ,\left\vert \Psi^{-}\right\rangle \right\}  $. The measurement outcomes $P_{1}$ and $P_{2}$ correspond
to
\begin{equation}
P_{\left\{ \left\vert \Phi^{+}\right\rangle ,\left\vert \Phi^{-}\right\rangle \right\} }=\left\vert \Phi^{+}\right\rangle
\left\langle \Phi^{+}\right\vert +\left\vert \Phi^{-}\right\rangle \left\langle \Phi ^{-}\right\vert
\end{equation}
and
\begin{equation}
P_{\left\{ \left\vert \Psi^{+}\right\rangle ,\left\vert \Psi^{-}\right\rangle \right\}  }=\left\vert \Psi^{+}\right\rangle
\left\langle \Psi^{+}\right\vert +\left\vert \Psi^{-}\right\rangle \left\langle \Psi^{-}\right\vert,
\end{equation}
respectively, where $\left\vert \Phi^{\pm}\right\rangle=(\left\vert 00\right\rangle \pm \left\vert 11\right\rangle)/\sqrt{2}$ and $\left\vert \Psi^{\pm}\right\rangle=(\left\vert 01\right\rangle \pm \left\vert 10\right\rangle)/\sqrt{2}$. More generally, one can obtain non-destructive projections onto subspaces spanned by two arbitrary Bell states
using additional single qubit unitary operations which allow one to permute Bell states. For instance, the application
$H\otimes H$ (here $H$ notes Hadamard gate) consequently before and after the measurement $P_{\left\{  \left\vert
\Phi^{+}\right\rangle ,\left\vert \Phi^{-}\right\rangle \right\}  }$ corresponds to $P_{\left\{  \left\vert
\Phi^{+}\right\rangle ,\left\vert \Psi^{+}\right\rangle \right\}  }$ because of
\begin{equation}
H\otimes H\ket{\Phi^+}=\ket{\Phi^+},H\otimes H\ket{\Phi^-}=\ket{\Psi^+}.
\end{equation}
Obviously, using these non-destructive projections, we can achieve a full BSM on atoms.

\section{entanglement purification of atomic states}
Entangled states are the most important resource for quantum information processing, including quantum teleportation
\cite{teleportation1,teleportation2}, efficient quantum computation \cite{computation} and quantum cryptography \cite{cryptography}.
Generally, these applications work best with pure, maximally entangled quantum states. However, due to dissipation and
decoherence, practically available states are likely to be non-maximally entangled, partially mixed, or both. To obtain useful
entangled states, there are various entanglement purification schemes proposed, which differ in their purification range, the
number of copies of the state they operate on. In this section, we focus on entanglement distillation which operate on a
single copy, recurrence protocol which operates on two copies at each step.

\subsection{Bipartite Entanglement distillation}

Entanglement distillation \cite{purification1,purification2,purification3} aims at preparing highly entangled states out of a supply of weakly entangled
pairs, using local devices and classical communication only. The simplest protocol operate on a single copy of an arbitrary
mixed state $\rho$ and consist in the application of a POVM. Hence sequences of local operations are applied in such a way
that for specific measurement outcomes the resulting state is more entangled than the initial state. Consider a special kind
of mixed states including certain rank two states of the form
\begin{equation}
\rho=F\ket{\Psi^+}\bra{\Psi^+}+(1-F)\ket{11}\bra{11},
\end{equation}
where $\ket{\Psi^+}=(\ket{01})+\ket{10}/\sqrt{2}$ is one of the Bell states. We apply two POVMs (local flittering operators)
\begin{equation}
A=B=\ket{0}\bra{0}+\varepsilon\ket{1}\bra{1}
\end{equation}
on the mixed state and obtain the resulting state (non-normalized)
\begin{equation}
\rho'=F\varepsilon^2\ket{\Psi^+}\bra{\Psi^+}+(1-F)\varepsilon^4\ket{11}\bra{11}.
\end{equation} The fidelity of the resulting
state is given by
\begin{equation}
F'=F\varepsilon^2/[F\varepsilon^2+(1-F)\varepsilon^4].
\end{equation}
 As expected, if
$\varepsilon$ is small enough, the improved fidelity $F'$ goes to 1 while the probability
\begin{equation}
p=F\varepsilon^2+(1-F)\varepsilon^4
\end{equation} to obtain the Bell state $\ket{\Psi^+}$ goes to zero. There is a tradeoff
between the reachable fidelity of the output state and the success probability.

Now we extend this method to any arbitrary bipartite mixed state and
investigate how its entanglement changes under local operations and
classical communications of the type $\rho'\sim (A\otimes
B)\rho(A\otimes B)^\dagger$. Consider a pure entangled state
$\ket{\varphi}=a\ket{00}+b\ket{11}$ (for simplicity, we assume $(a
\ge b) \in \mathbb{R}$) evolving through the same locally
depolarizing channel $M_i\in
{\{\sqrt{1-p}\openone,\sqrt{p/3}\sigma_x,\sqrt{p/3}\sigma_y,\sqrt{p/3}\sigma_z\}}$
respectively, which is one of the most typical noisy channels in
atomic systems. The resulting mixed state reads
\begin{widetext}
\begin{align}
&\rho=\sum_{i,j}(M_i\otimes M_j)\ket{\varphi}\bra{\varphi}(M_i\otimes M_j)^\dagger \\
&=\frac{1}{9}\left(\begin{array}{cccc} 3a^2(3-4p)+4p^2 & 0 & 0 & ab(3-4p)^2 \\ 0 & 6p-4p^2 & 0 & 0 \\ 0 & 0 & 6p-4p^2 & 0
\\ ab(3-4p)^2 & 0 & 0 & 3b^2(3-4p)+4p^2\end{array}\right). \notag
\end{align}
\end{widetext}

The optimal local filtering operations for a single copy of mixed
state has been obtained by Verstraete \emph{et.al.} in
\cite{purification1,purification2,purification3}. Following their discussions, we introduce the
real and linear parametrization of the $\rho$ (R-picture) as
\begin{eqnarray}
\rho &=& \frac{1}{4} \sum_{i,j} R_{ij} \sigma_i \otimes \sigma_j,
\\
R_{ij} &=& \text{Tr}[\rho(\sigma_i\otimes \sigma_j)],
\end{eqnarray}
with $\sigma_0=\openone$ and $\sigma_1$,$\sigma_2$,$\sigma_3$ are
the usual Pauli matrices. By using the singular value decomposition
in Lorentz metric, the $4\times 4$ matrix $R$ can be decomposed as
$R = L_1 \Sigma L_2^T$, where $L_1$, $L_2$ are some proper
orthochronous Lorentz transformations, and  $\Sigma$ is of some
normal form depending on $R$. The concurrence and the density matrix
after the distillation are totally determined by $\Sigma$. In this
case, we have $\Sigma=\mbox{diag}\{s_0, s_1, s_2, s_3\}$ with
\begin{align}
s_0 &= \frac{2}{9}( -4 p^2+6 p+\Delta), \notag \\
s_1 &= s_2=\frac{2}{9}a b (3-4 p)^2, \notag \\
s_3 &= -\frac{2}{9}(4 p^2-6 p+\Delta), \notag \\
\Delta &= \sqrt{9a^2b^2(3-4p)+4p^2(3-2p)^2}.
\end{align}

The optimal local distillation operations can be easily obtained
from $L_1$ and $L_2$ directly and given by
\begin{eqnarray}
A\otimes A^* = T^{\dag} M L_1^T M T, \\
B\otimes B^* = T^{\dag} M L_2^T M T, \\
T=\frac{1}{\sqrt{2}}\left(\begin{array}{cccc}
1&0&0&1\\0&1&1&0\\0&\mathrm{i}&-\mathrm{i}&0\\1&0&0&-1\end{array}
\right),
\end{eqnarray}
with $M$ the usual Lorentz metric matrix $M=\mbox{diag}\{1,
-1,-1,-1\}$. After a straightforward algebra, the optimal POVM can
be expressed as follows:
\begin{equation}
A\otimes B= \left(\begin{array}{cc} 0 & 1 \\ \mathrm{i}\left|\frac{1-c}{1+c}\right|^{1/2} & 0 \end{array}\right)\otimes
\left(\begin{array}{cc} -\mathrm{i}\left|\frac{1-c}{1+c}\right|^{1/2} & 0 \\0 & 1 \end{array}\right),
\end{equation}
where
\begin{equation}
c=\frac{8p^2-12p+9+2\Delta}{3(a^2-b^2)(3-4p)}.
\end{equation}
$B$ can be implemented with the setup in Fig.\ \ref{fig1} by adding a phase shifter (QWP) behind partial polarizer, while $A$ is
obtained by a POVM in Eq. (2.10) followed by a single qubit rotation $\sigma_x$ on the atomic state by choosing the proper
$\varepsilon=\left|\frac{1-c}{1+c}\right|^{1/2}$.

The output state obtained after the application of local filtering
operations can be represented as
\begin{equation}
\rho'=\frac{\left(A \otimes B\right)\rho \left(A \otimes B\right)^\dagger}{\text{Tr}\left[\left(A \otimes B\right)\rho \left(A
\otimes B\right)^\dagger\right]}= \frac{1}{4} \sum_{i,j} R'_{ij} \sigma_i \otimes \sigma_j,
\end{equation}
with $R'=\Sigma/s_0$. The distillation protocol produces the state
$\rho'$ with maximal possible entanglement of formation (EoF)
according to the concurrence as introduced by Wootters
\cite{Wooters89}
\begin{align}
C(\rho')&=\max\{0,\frac{-s_0+s_1+s_2-s_3}{2s_0}\}\notag\\
&=\max\{0, \frac{ab(3-4p)^2+4p^2-6p}{-4p^2+6p+\Delta}\},
\end{align} compared to that of the initial state
\begin{align}
C(\rho)&=
\max\{0,\frac{-s_0+s_1+s_2-s_3}{2}\}\notag\\&=\max\{0,\frac{2}{9}[ab(3-4p)^2+4p^2-6p]\}.
\end{align}
Additionally the output state also has the maximum possible
violation of the Clauser-Horne-Shimony-Holt (CHSH) version of
inequality. Consider the expectation value
$\beta_{\rho}=\text{Tr}(\rho\mathcal{B})$ for the operator
\begin{equation}
\mathcal{B}=\frac{1}{2}\sum_{i,j=1}^3\left[a_i(c_j+d_j)+b_i(c_j-d_j)\right]\sigma_i\otimes\sigma_j,
\end{equation}
where $(\vec{a}, \vec{b}, \vec{c}, \vec{d})$ are real unit vectors.
In the CHSH inequality, the proposed value $\beta_{\rho}$, a
combination of four polarization correlation probabilities, should
not be more than 1 for local hidden variables theory (to obtain the
value of the usual CHSH inequality, the factor $2$ should be
involved). In our case, the maximal CHSH violations for both initial
and purified states are shown as
\begin{align}
&\beta_\rho=\frac{2\sqrt{2}}{9}ab(3-4p)^2,\notag \\
&\beta_{\rho'}=\frac{\sqrt{2}ab(3-4p)^2}{-4p^2+6p+\Delta}.
\end{align}

In this paper we will focus on how to implement the entanglement
distillation on neutral atoms, so next another typical noise
channel---amplitude damping channel is considered. The corresponding
super-operators are $M_i\in\{\left(\begin{array}{cc}1 & 0
\\0 & \sqrt{1-p}\end{array}\right),\left(\begin{array}{cc}0 & \sqrt{p}\\0 & 0 \end{array}\right)\}$. An entangled
state $\ket{\varphi}=a\ket{00}+b\ket{11}$ passes through the noise channel and then we obtain a mixed state
\begin{equation}
\rho = \left(\begin{array}{cccc} a^2+b^2p^2 & 0 & 0 & ab(1-p)\\ 0& b^2(1-p)p & 0 & 0 \\ 0 & 0 & b^2(1-p)p & 0
\\ab(1-p) & 0 & 0 & b^2(1-p)p \end{array}\right).
\end{equation}

If the initial state is one of Bell state $\ket{\Phi^+}$, i.e. $a=b=1/\sqrt{2}$, the $4\times 4$ matrix $R$ is shown as
\begin{eqnarray}
R &=& \left(
\begin{array}{llll}
 1 & 0 & 0 & p  \\
 0 & 1-p  & 0 & 0 \\
 0 & 0 & p-1 & 0 \\
 p  & 0 & 0 & 2p^2-2p+1
\end{array}
\right).
\end{eqnarray}

The optimal local filtering operation can be simplifies as
\begin{equation}
A\otimes B= \left(\begin{array}{cc}\left|\frac{1-p}{\sqrt{p ^2+1}}\right|^{1/2} & 0\\0 & 1\end{array}\right)\otimes \left(\begin{array}{cc}0 & 1\\
\left|\frac{1-p}{\sqrt{p ^2+1}}\right|^{1/2} & 0\end{array}\right).
\end{equation}

Similarly, the POVMs $A$ and $B$ can be implemented by cavity-assistant interaction and linear optical elements with the
proper choices of $\varepsilon=\left|\frac{1-p}{\sqrt{1+p^2}}\right|^{1/2}$.

After the application of local filtering operations, the $4\times4$ matrix $R'$ according to the output state $\rho'$ is
obtained as
\begin{eqnarray}
R' &=& \frac{\sqrt{p ^2+1} - p}{1-p} \Sigma \notag \\
&=& \left( \begin{array}{cccc} 1 & 0 & 0 & 0\\ 0 & \sqrt{p ^2+1} - p & 0 & 0\\ 0 & 0 & \sqrt{p ^2+1} - p & 0 \\
0 & 0 & 0 & -(\sqrt{p ^2+1} - p)^2 \end{array} \right)
\end{eqnarray}
with the concurrence
\begin{equation}
C(\rho')=\max\{0,\frac{1-p}{\sqrt{p^2+1}+p}\}
\end{equation}
compared to that of the initial mixed state $C(\rho)=\max\{0,(1-p)^2\}$. The maximal CHSH violations for both initial and
purified states are obtained as
\begin{equation}
\beta_\rho=\sqrt{2}(1-p), \beta_{\rho'}=\frac{\sqrt{2}}{\sqrt{p^2+1}+p}.
\end{equation}

Next we distill the entanglement from the general two-qubit mixed state. We let two qubits of the nonmaximally entangled
states pass through the same phase damping channel in basis
$\{\ket{0},\ket{1}\}$, respectively. The corresponding
super-operators are
$\{\sqrt{1-p}\openone,\sqrt{p}\sigma_\text{z}\}_\text{A}\otimes
\{\sqrt{1-p}\openone,\sqrt{p}\sigma_\text{z}\}_\text{B}$. After
decoherence, we obtain a mixed state shown as
\begin{equation}
\rho=\left(\begin{array}{cccc} a^2 & 0 & 0 & ab(-1+2p)^2\\ 0& 0 & 0 & 0 \\ 0 & 0 & 0 & 0
\\ab(-1+2p)^2 & 0 & 0 & b^2 \end{array}\right).
\end{equation}
For this kind of mixed states, the only local operation is an identity on the first particle and a POVM (unilateral local
filtering) on the second particle $A\otimes B=\openone\otimes \left(\begin{array}{cc} b & 0\\ 0 & a
\end{array}\right)$. The POVM $B$ can be implemented by cavity-assistant interaction and linear optical elements with the
proper choices of $\varepsilon=a/b$. After the application of local filtering operations, the output state is obtained as
$\rho'\propto (A\otimes B)\rho(A\otimes B)^\dagger$ and in R-picture, the matrix R' can be decomposed as
\begin{equation}
R' =\left( \begin{array}{cccc} 1 & 0 & 0 & 0\\ 0 & 1 & 0 & 0\\ 0 & 0 & (1-2p)^2  & 0 \\
0 & 0 & 0 & -(1-2p)^2 \end{array} \right).
\end{equation}
The corresponding concurrences and CHSH violations can also be
obtained and list as follows
\begin{eqnarray}
C(\rho')&=& \frac{1}{2 |ab|} C(\rho) = (1-2 p)^2, \nonumber \\
\beta_{\rho'} &=& \frac{1}{2 |ab|} \beta_{\rho} = \sqrt{1+(1-2 p)^4}.
\end{eqnarray}

The entanglement distillation protocol can be generalized to arbitrary two-qubit partially mixed state, since any nontrivial
local operation and classical communication can be written as the form $\gamma U_\text{A}\left(\begin{array}{cc} 1 & 0\\ 0 &
\alpha
\end{array}\right)U'_\text{A}\otimes U_\text{B}\left(\begin{array}{cc} 1 & 0\\ 0 &
\beta
\end{array}\right)U'_\text{B}$, where $U_\text{A(B)}$ and $U'_\text{A(B)}$
denote local unitary operations on the first (second) and $\gamma$ is a scale factor in the range $0\le\gamma\le1$ and $0\le
\alpha (\beta)\le 1$.

\begin{figure}[ht]
  \includegraphics[width=9 cm]{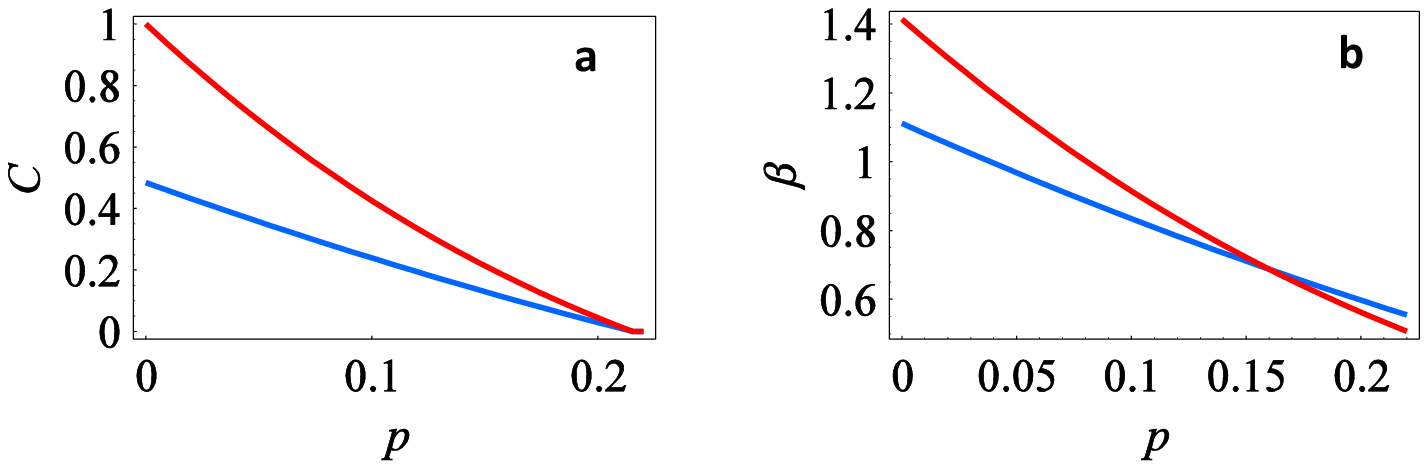}
   \caption{The concurrence $C$ (a) and violation of CHSH inequality $\beta$ (b) along with the parameter of the depolarizing channel $p$.
   Here we show an example with the initial state $0.97\ket{00}+0.25\ket{11}$. The blue and red lines are for the mixed state before and
   after the distillation process.}
   \label{fig2}
\end{figure}

\begin{figure}[ht]
  \includegraphics[width=9 cm]{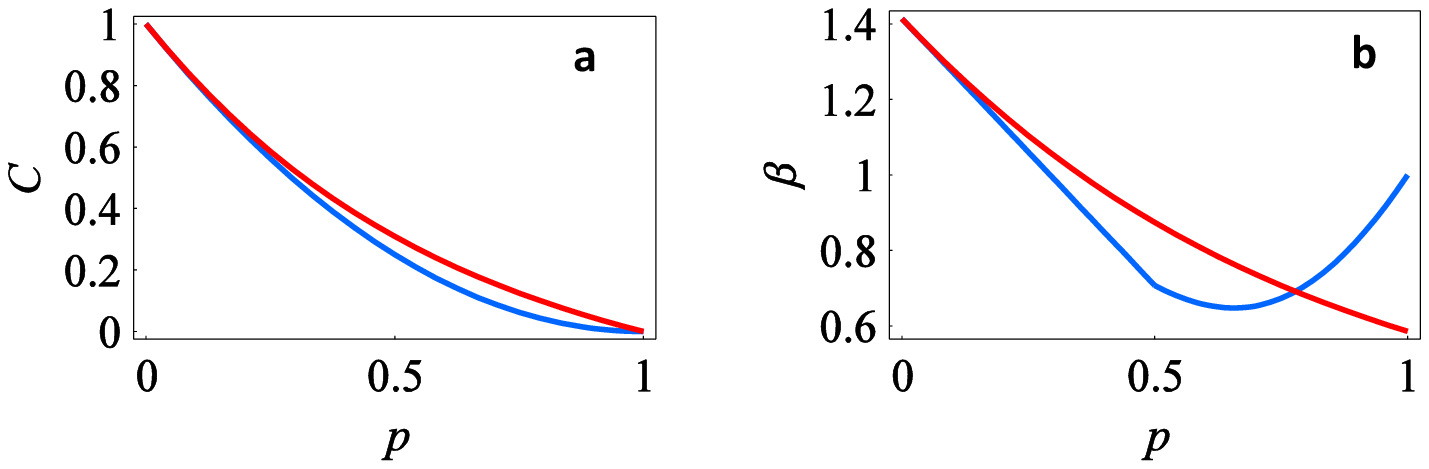}
   \caption{The concurrence $C$ (a) and violation of CHSH inequality $\beta$ (b) along with the parameter of the damping channel $p$.
   Here we show an example with the initial state $\left(\ket{00}+\ket{11}\right)/\sqrt{2}$.
   The blue and red lines are for the mixed state before and after the distillation process.}
   \label{fig3}
\end{figure}

Figs. \ \ref{fig2} and \ \ref{fig3} show the concurrences and violations of CHSH inequality
along with the parameter $p$ for depolarizing channel and amplitude
damping channel respectively. One can find that local filtering
operations can always yield a new state with larger entanglement of
formation. And there exist some cases where the initial mixed states
do not violate the CHSH inequality while violate them after the
local filtering operations. This also compatible with the results in
\cite{purification1,purification2,purification3}. However, we can also find that although
$C(\rho)$ and $\beta_{\rho}$ are both a entanglement measure, they
are not just equal. Local filtering operations do not always lead to
a larger violation of the CHSH inequality. In Fig.\ \ref{fig2}, if $p$ is
smaller than 0.16, the concurrence and violation of CHSH inequality
of the state after the distillation process are greater than those
of the mixed state before the distillation; if $p$ is greater than
0.16, the concurrence of the state after the distillation is still
greater than that of the initial mixed state, while the violation of
the state after the distillation is even smaller. For the damping
channel, it is similar and the critical point is $p\approx 0.775$.
Thus the concurrence and violation of the CHSH inequality are two
independent entanglement measures.

\subsection{recurrence protocols}

In the following we discuss another protocol to produce states arbitrarily close to a maximally entangled pure state by
iterative application. That is so-called the recurrence protocol.

Consider a mixed atomic state $\rho_{12}$ resulting from an imperfect distribution of $\left\vert \Phi^{+}\right\rangle $. We
decompose $\rho$ into two terms, $\rho=\widehat{\rho}+\rho_{od}$, and express the density operator $\widehat{\rho}$ in the
Bell basis $\left\{\left\vert \Phi^{+}\right\rangle ,\left\vert \Phi^{-}\right\rangle ,\left\vert \Psi^{+}\right\rangle
,\left\vert \Psi^{-}\right\rangle \right\}$ and denote the diagonal elements in that basis by $\left\{ A,B,C,D\right\}$. All
off-diagonal elements in the Bell basis ($\rho_{od}$) is made irrelevant by the protocol. Note that the first diagonal element
$A=\left\langle \Phi^{+}\right\vert \rho\left\vert \Phi^{+}\right\rangle $, which is the definition of the fidelity.

Given two mixed states $\rho_{12}$ and $\rho_{1^{\prime}2^{\prime}}$, described by $\left\{  A,B,C,D\right\}$ and $\left\{
A^{\prime},B^{\prime },C^{\prime},D^{\prime}\right\}$ (and the irrelevant $\rho_{od}$) respectively, the following sequence of
local operations obtains with a certain probability a state with higher fidelity and hence purifies the state: (i) application
of $\sigma_{x,1}\otimes\sigma_{x,2}$ or $\openone$ with probability 1/2 on $\rho_{12}$ and similar to
$\rho_{1^{\prime}2^{\prime}}$; (ii) the partial BSM on both $\rho_{12}$ and $\rho_{1^{\prime}2^{\prime}}$, i.e.,
$\sum_{i=1,2}\sum_{j=1,2}M_{i}^{11^{\prime}}M_{j}^{22^{\prime}}\rho_{12}\rho_{1^{\prime}2^{\prime}}$, where
$M_{1}=P_{\left\{\left\vert \Phi^{+}\right\rangle ,\left\vert \Phi^{-}\right\rangle \right\}}$ and $M_{2}=P_{\left\{\left\vert
\Psi^{+}\right\rangle ,\left\vert \Psi^{-}\right\rangle\right\}  }$, which can be implemented with projection measurements shown in Sec II.
We only keep the state $\rho_{12}$ if $i=j$, i.e. the results in the final measurement coincide in both states.

The effect of (i) is to erase off-diagonal terms of the form $\left\vert \Phi^{\pm}\right\rangle\!\left\langle
\Psi^{\pm}\right\vert $ which may contribute to the protocol. The operation in (ii) is to select states in atoms 1' and 2'
which are eigenstates of $\sigma_z \otimes \sigma_x$ with eigenvalue +1 while eigenstates with eigenvalue -1 are discarded. In
particular, we find that the remaining off-diagonal elements do not contribute and the action of the protocol can be described
by the non-linear mapping of corresponding vector $\overrightarrow{x}=(A,B,C,D)$,
$\overrightarrow{x}^{\prime}=(A^{\prime},B^{\prime},C^{\prime},D^{\prime})$. The resulting state is of the form
$\tilde{\hat{\rho}}+\tilde{\rho}_{od}$, and $\tilde{\hat{\rho}}$ will on average have diagonal elements given by
\begin{align}
\widetilde{A}&=\left(AA^{\prime}+CC^{\prime}\right)  /N,\widetilde{B}=\left(BB^{\prime}+DD^{\prime}\right) /N,\\
\widetilde{C}&=\left(BD^{\prime}+DB^{\prime}\right) /N, \widetilde{D}=\left(  AC^{\prime}+CA^{\prime}\right) /N,\nonumber
\end{align}
and $N=\left(A+D\right)\left(A^{\prime}+D^{\prime}\right)+\left(B+C\right)  \left(B^{\prime}+C^{\prime}\right)$ is the
probability of success of the protocol. This map is equivalent to the purification map obtained in Ref.\ \cite{Deutsch} for
non-encoded Bell states. It follows that an iteration of the map--which corresponds to iteratively applying the purification
procedure (i-ii) to two identical copies of states resulting from successful previous purification rounds--leads to a
maximally entangled state. That is, the map has $\left\{1,0,0,0\right\}$ as attracting fixed point whenever $A>B+C+D$. We
emphasize that all errors leading outside the subspace, independent of their probability of occurrence, can be corrected. The
method is capable of purifying a collection of pairs in any state $\widehat{\rho},$ whose average fidelity with respect to at
least one maximally entangled state with probability greater than $1/2$. The central part of recurrence protocol of
entanglement purification---partial BSM can be implemented on neutral atoms via cavity-assistant interaction and linear
optical elements shown above.

Here we remark that the methods such as nested purification can also be applied via BSMs, which significantly reduced the
required number nodes \cite{Wolfgang}.

\section{Feasibility}

In this section, we analyze the feasibility of this scheme using numerical simulations in the experimental parameters.

No particularly demanding assumptions have been made for the experimental parameters. The relevant cavity QED parameters for
our system are assumed to be $\left(g_0,\kappa,\gamma\right) /2\pi=\left(27,2.4,2.6\right)$ MHz \cite{Sauer}.
$g_0^2/\kappa\gamma=117\gg1$ places our system well into the strongly coupled regime. The cavity consists of two $1$-mm-diam
mirrors with $10$ cm radii of curvature separated by $75$ $\mu$m \cite{Sauer} assuming the wavelength of the cavity mode is
$\sim780$ nm (the rubidium $D2$ line). The distance between two atoms $d$ in an optical lattice is about $10$ $\mu$m, which is
larger than the waist of the cavity $\sim5$ $\mu$m, leaving only one atom inside the cavity and its neighbor atoms outside for
the gate operations.

The evolution of the two-atom states is accomplished during the passing time of the single-photon pulse $T\sim200/\kappa=10$
$\mu$s. The maximum velocity of the atoms in the transverse optical lattices and the maximum acceleration imparted are about
$30$ cm/s and $1.5$g, respectively. Moving the proper atoms into and outside of the cavity is accomplished within a time
$\tau_{\mathrm{T}}\approx 100$ $\mu$s. The gate preformation and the transport of atoms can be accomplished within the
coherent time (dephasing) of the atoms, lasting milliseconds or hundreds of milliseconds, which depends on the sensitivity to
the magnetic fluctuations of the internal atomic states \cite{Ja,Bonn1,Bonn2,Bonn3}.

The partial polarizer has been implemented experimentally. For example, the partial polarizer with four slabs \cite{Kwiat}
allows for a transmission probability $T_{\mathrm{h}}=0.18$ (horizontal polarization) and $T_{\mathrm{v}}=0.89$ (vertical
polarization) respectively. The transmission of horizontally polarized light is thus only $20\%$ of that for the vertically
polarized light, i.e. $\varepsilon=0.2$. In principle, by tuning $\theta$, the angle between the partial polarizer and the
vertical axis, we can obtain any value of $\varepsilon$ (see Appendix for details). Hence, our scheme fits well the status of
current experimental technology.

Experimentally, major sources of deviations from the ideal setup are found to be: addressing errors, spontaneous emissions and
long term interferometric phase instability. Since we only apply local operations, there is no interferometer required in the
setup of our proposal. Also, spontaneous emissions only lead to photon losses which can be detected, merely decrease the
probability of success and thus have no contribution to a lack of fidelity.

\begin{figure}[hb]
   \includegraphics[width=.47\textwidth]{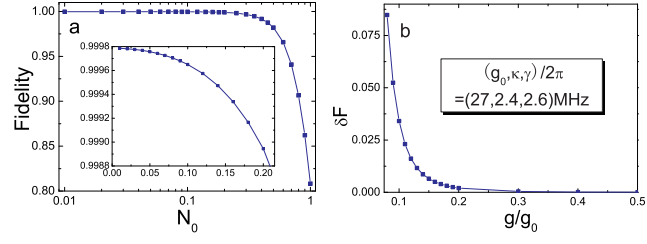}
   \caption{(a) The fidelity of the POVM shown in Eq. (2.10) versus $g/2\pi$ with pulse duration $T=10\mu$s, $\kappa/2\pi=2.4$MHz,
   $\gamma/2\pi = 2.6$MHz, $\varepsilon=0.2$. We have assumed a Gaussian shape for the input pulse
   with $f(t)\propto\exp\left[-\left(t-T/2\right)^2/\left(T/5\right)^2\right]$. (b) It changes with $g/g_{o}$.}
   \label{fig4}
\end{figure}

Some additional sources of fidelity degradation have been estimated by numerical simulations, which show that the proposed
scheme works remarkably well. In these simulations we considered the fidelity of the POVM shown in Fig.\ \ref{fig4}, which is the same to the fidelity
of the output states $\rho'$ obtained after entanglement distillation. Further decrease of fidelity is for example caused by
shape mismatching between the input and output pulses. Fig.\ \ref{fig2}a shows that we retain a high fidelity of the POVM
under $\left(\kappa,\gamma\right) /2\pi=\left(2.4,2.6\right)$ MHz (for all atoms with equal coupling coefficient $g$) and
$\varepsilon=0.2$ $-$which means that $\mathcal{F}_\text{POVM}$ is up to $99\%$ for $g/2\pi>27$MHz \cite{Sauer}. Our method is
also insensitive to randomness in the coupling rates caused by fluctuations in the position of the atom, which can be
numerically calculated by variation of $g$. As the figure shows, $\mathcal{F}_\text{POVM}$ is extremely insensitive to this
influence, as $\delta \mathcal{F}_\text{POVM}$, describing the change of the fidelity, stays below $10^{-3}$ for $g$ varying
from $27$ MHz to $13.5$ MHz. It is shown in Fig.\ \ref{fig2} that the fidelity of POVM.

From Sec. II (C), the fidelity of the operation of BSM is about $\mathcal{F}_\text{BSM}\approx \mathcal{F}^2_\text{POVM}$. The
fidelity of the final state obtained after purification with the recurrence protocol is then $\tilde{A}
\mathcal{F}_\text{BSM}$.

The efficiency of this scheme is characterized through exact numerical simulations that incorporate various sources of noise
and this demonstrates the practicality with the background of current experimental technology.

\section{Conclusion}

In summary, we present a scheme to realize bipartite entanglement purification on neutral atoms via cavity-assistant interaction
and linear optical elements. Two kinds of purification protocols are proposed here. That is the entanglement distillation
which can be realized by POVM on atomic states and the recurrence protocol based on partial BSM. The proposal has been tested by
numerical simulations including various sources of noise present in experiments, and was found to be feasible for experimental
realization based on the status of current technology.

\begin{appendix}
\begin{center}\bf{Appendix}\end{center}

\begin{figure}[ht]
   \includegraphics[width=.45\textwidth]{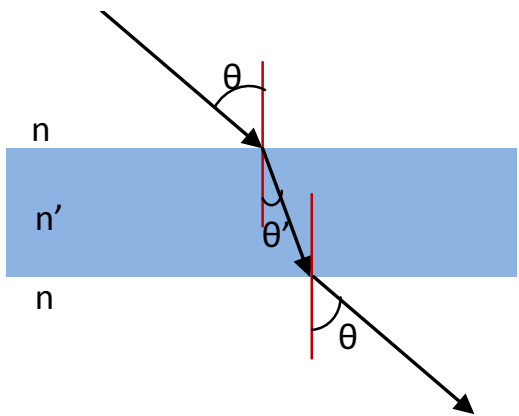}
   \caption{The relationship between the transmission probabilities and the angle $\theta$ can be always established based on Fresnel equation.}
   \label{appendix}
\end{figure}

In principle, we can always establish a relationship (shown in Fig.\ \ref{appendix}) between the transmission probability and the angle $\theta$ between the
partial polarizer and the vertical axis based on Fresnel equation. At the upper interface, we obtain the transmission
probabilities $t_\text{h}$ and $t_\text{v}$ as
\begin{align}
&t_\text{h}=\frac{2n \text{cos}\theta}{n' \text{cos}\theta+n \text{cos}\theta'}=\frac{2 \text{cos}\theta
\text{sin}\theta'}{\text{sin}(\theta+\theta')\text{
cos}(\theta-\theta')};\tag{A1.a} \\
&t_\text{v}=\frac{2n \text{cos}\theta}{n \text{cos}\theta+n'
\text{cos}\theta'}=\frac{2\text{cos}\theta\text{sin}\theta'}{\text{sin}(\theta+\theta')} \tag{A1.b},
\end{align}
where $n$ and $n'$ are the refractive indexes. Consider both interfaces, we obtain the total transmission probabilities as
\begin{align}
&T_\text{h}=t^{(1)}_\text{h}t^{(2)}_\text{h}=\frac{\text{sin}2\theta\text{sin}2\theta'}{\text{sin}^2(\theta+\theta')\text{cos}^2(\theta-\theta')};
\tag{A2.a} \\
&T_\text{v}=t^{(1)}_\text{v}t^{(2)}_\text{v}=\frac{\text{sin}2\theta\text{sin}2\theta'}{\text{sin}^2(\theta+\theta')}
\tag{A2.b}.
\end{align}
If substituting Eq.~(A2.a) into (A2.b), we will obtain $\varepsilon=T_\text{h}/T_\text{v}$ as functions of $\theta$.
\end{appendix}

\begin{acknowledgments}
This work was founded by NSERC, MITACS, iCORE, QuantumWorks, CIFAR, the National Fundamental Research Program (2006CB921900),
the National Natural Science Foundation of China (Grant No. 10674127), the Innovation Funds from the Chinese Academy of
Sciences, and Program for New Century Excellent Talents in University.
\end{acknowledgments}

\newpage

\newpage
\newpage
\newpage
\newpage

\end{document}